\documentclass{article}
\usepackage[T1]{fontenc} 
\usepackage[utf8]{inputenc} 
\usepackage{ismir,amsmath,cite,url}
\usepackage{graphicx}
\usepackage[bookmarks=false]{hyperref}

\usepackage{color}

\def\lbd{}	

\usepackage{lineno}

\title{Symbotunes: unified hub for symbolic music generative models}

\multauthor
{Paweł Skierś$^{*}$ \hspace{1cm} Maksymilian Łazarski$^{*}$ \hspace{1cm} Michał Kopeć$^{*}$ \hspace{1cm} Mateusz Modrzejewski}
{Institute of Computer Science, Warsaw University of Technology, Poland\\
{\tt\small pawel.skiers.stud@pw.edu.pl, maksymilian.lazarski.stud@pw.edu.pl,}\\
{\tt\small michal.kopec6.stud@pw.edu.pl, mateusz.modrzejewski@pw.edu.pl}\\
$^*$ - equal contribution
} 

\def\authorname{P. Skierś, M. Łazarski,  M. Kopeć and M. Modrzejewski}

\begin{document}

\maketitle
\begin{abstract}
Implementations of popular symbolic music generative models often differ significantly in terms of the libraries utilized and overall project structure. Therefore, directly comparing the methods or becoming acquainted with them may present challenges. To mitigate this issue we introduce Symbotunes, an open-source unified hub for symbolic music generative models. Symbotunes contains modern Python implementations of well-known methods for symbolic music generation, as well as a unified pipeline for generating and training. 
\end{abstract}
\section{Introduction}\label{sec:introduction}

Symbolic music models have been instrumental in advancing research on music analysis and generation, but many of the foundational models were developed using now-obsolete frameworks, such as Theano, which hinders reproducibility and continued experimentation. This also creates difficulties and barriers for researchers new to the field. We introduce Symbotunes, a unified hub for symbolic models. Symbotunes addresses the aforementioned issues by offering a framework that re-implements these symbolic models, increasing compatibility with modern machine learning practices. Designed for both researchers and educators, Symbotunes provides a standardized platform that facilitates the exploration, adaptation, and further development of symbolic music models.

\section{Design choices}\label{sec:design_choices}
To manage our hub's required dependencies we use anaconda\cite{anaconda} package manager. We provide an \texttt{environment.yml} file to facilitate a reproducible environment. Notably, in our hub, we use Python 3.12, PyTorch 2.2 \cite{paszke2017automatic}, PyTorch Lightning 2.2 \cite{falcon2019pytorch} and MidiTok 3.0 \cite{fradet2023miditok}.

\subsection{Project structure}
The project is divided into four main subfolders:
\begin{itemize}
    \item \textbf{scripts} - contains Python scripts for training and sampling from the models,
    \item \textbf{models} - contains the models currently implemented in our hub. Each model is in a separate sub-directory, which also contains an example training script,
    \item \textbf{data} - contains all the data handling utilities available in the hub - datasets, tokenizers, and data transforms. It provides a simple, unified interface for symbolic music datasets,
    \item \textbf{callbacks} - contains the training callbacks.
\end{itemize}

\subsection{Configuration system}
One of the key parts of Symbotunes is our custom configuration file parser. It allows the user to run different experiments with a single Python script, by simply changing the configuration file. Notably, the configs allow the users to specify the model type and its hyperparameters, datasets, and data transformations such as tokenizers, training callbacks, and training details such as the number of training steps. For the config file structure, we use yaml, due to the flexibility and readability of the format.

\subsection{Implemented models}
All the available models inherit from the abstract \texttt{BaseModel} class, ensuring a common interface. Currently, Symbotunes contains implementations of the following symbolic music methods:
\begin{itemize}
    \item Folk-RNN \cite{sturm2016music} - an LSTM-based model designed to generate traditional folk music in ABC notation based on training data from a large collection of folk tunes,
    \item MusicVAE \cite{roberts2018hierarchical} - a model developed by Google’s Magenta team that uses a recurrent variational autoencoder (VAE) to generate and interpolate between musical sequences,
    \item ABC GPT2 \cite{geerlings2020interacting} - model utilizing the transformer architecture to generate ABC samples.
\end{itemize}

\subsection{Datasets and data handling}
To ensure interface unification, all the available datasets inherit from the abstract \texttt{BaseDataset} class. Currently, we provide the following music datasets:
\begin{itemize}
    \item \textbf{LAKH dataset} \cite{raffel2016learning} - dataset containing popular songs in MIDI format,
    \item \textbf{Folk RNN dataset} \cite{sturm2016music} - contains folk music in ABC format retrieved for the Folk-RNN model.
\end{itemize}
We also provide a set of useful data transforms compatible with the datasets:
\begin{itemize}
    \item \textbf{folk-rnn tokenizer} - tokenizer for ABC format as described in \cite{sturm2016music}, takes string as input and outputs tokenized sequence,
    \item \textbf{midi tokenizer} - generic MidiTok REMI MIDI file tokenizer, takes path to MIDI file as input, outputs MidiTok \texttt{TokSequence},
    \item \textbf{sample bars} - samples $n$ random bars (continuous, one bar after another) from the given track. Takes \texttt{TokSequence} as input and outputs $n$ sampled bars or the original sequence if the track has less than $n$ bars (as \texttt{TokSequence}),
    \item \textbf{toksequence to tensor} - transforms \texttt{TokSequence} to PyTorch \texttt{Tensor},
    \item \textbf{sample subsequence} - samples sub-sequence of given length from the given sequence (or return the sequence if the sequence is shorter than the expected sub-sequence).

\end{itemize}
\subsection{Experiment logging}
Symbotunes provides a convenient way of logging the experiment results, by utilizing the Weights and Biases \cite{wandb} platform. Additionally, it also contains the following training callbacks:
\begin{itemize}
    \item \textbf{setup callback} - sets up log directory for the experiment. Creates a directory for model checkpoints, a directory with the experiment configuration, and saves the experiment configuration,
    \item \textbf{model checkpoint} - saves to the checkpoints the weights of the best model and the last model,
    \item \textbf{checkpoint every $n$ steps} - creates a model checkpoint every $n$ training steps,
    \item \textbf{CUDA callback} - after the first epoch, computes the maximum GPU memory usage during training.
\end{itemize}

\section{Usage}\label{sec:usage}

\subsection{Run scripts}
Symbotunes provides two main functionalities: training models from scratch and generating samples with trained models.
\begin{enumerate}
    \item \textbf{Training} - By running the \texttt{train.py} script users can train a model from scratch. The run specification is defined in the config file passed to the script via the \texttt{--path} flag. It is also possible to start the training from a pretrained checkpoint by using the \texttt{--checkpoint} flag.
    \item \textbf{Generating samples} - Users can sample from a trained model by running the \texttt{sample.py} script. It requires a model config file, a compatible checkpoint, and a number of samples to generate, which can be specified with \texttt{--path}, \texttt{--checkpoint}, and \texttt{--batch} flags respectively. The user can also optionally provide the output path to which the samples will be saved with \texttt{--out} flag.
\end{enumerate}
For more information and examples of run commands, we refer the reader to the hub's \texttt{README.md} file.

\subsection{Modifying the config files}
Each config file consists of four main sections:
\begin{enumerate}
    \item \textbf{model}: it specifies the model type and values of all of its hyperparameters,
    \item \textbf{dataloaders}: defines train and validation dataloaders used during the training,
    \item \textbf{lightning}: defines general training parameters such as the number of steps,
    \item \textbf{callbacks}: defines the callbacks that will be used during training.
\end{enumerate}
For each of the models available in the hub, we provide an example configuration file. Additionally, we provide a more detailed description of the file structure in the hub's \texttt{README.md}.

\section{License}\label{sec:license}
 We have opted to release Symbotunes under the GPL license. The source
 code is publicly accessible via GitHub at \href{https://github.com/pskiers/Symbotunes}{https://github.com/pskiers/Symbotunes}. We actively encourage community involvement and contributions either by submitting pull requests or by forking the repository.

\section{Conclusions}\label{sec:conclusions}
We present Symbotunes, a Python-based, open-source hub designed to solve compatibility and reproducibility challenges in symbolic music generative models. By reimplementing foundational models with PyTorch Lightning, Symbotunes offers a modern, standardized platform for researchers and educators to explore, adapt, and develop symbolic music models. Released under the GPL license, it invites open collaboration.

Future work includes adding more models, datasets, tokenizers, and features like model sample logging. Symbotunes aims to become a key resource for advancing symbolic music generation research and education.

\bibliography{ISMIRtemplate}

\end{document}